\documentclass[pra,aps,showpacs,floatfix,floats,superscriptaddress,nobalancelastpage,twocolumn]{revtex4-1}
\usepackage{graphicx}
\usepackage{graphics}
\usepackage{latexsym,amsmath,amssymb,bm,euscript}
\usepackage{color}
\usepackage{dcolumn}
\usepackage{bm}
\usepackage{epstopdf}
\usepackage{times}
\usepackage{multirow}
\usepackage{wasysym}
\usepackage{subfigure}
\usepackage{bbm}

\def\ra{\rangle}
\def\la{\langle}

\begin{document}

\title{Relaxation of a Goldstino-like mode due to supersymmetry breaking in Bose-Fermi mixtures}
\author{Hsin-Hua Lai}
\affiliation{National High Magnetic Field Laboratory, Florida State University, Tallahassee, Florida 32310, USA}
\author{Kun Yang}
\affiliation{Department of Physics and National High Magnetic Field Laboratory, Florida State University, Tallahassee, Florida 32306, USA}
\date{\today}

\begin{abstract}
In the presence of \textit{nonrelativistic} supersymmetry, a sharp fermionic collective mode similar to the Goldstino mode in high-energy physics was proposed to be realized in Bose-Fermi mixtures. The Goldstino mode is relaxed (a.k.a. decays) if supersymmetry is explicitly broken, which can be revealed as the broadening of the corresponding spectral function. We find that the situation shares many similarities with the electron spin resonance in magnetic systems and adopt the well-known Kubo-Tomita theory to perform a general analysis of the spectral function lineshape broadening of the Goldstino mode. 
\end{abstract}
\maketitle

\section{Introduction}
Supersymmetry (SUSY), the intriguing symmetry relating bosons and fermions, has been of strong interest in elementary particle physics after the first theoretical model in interacting quantum field theory with supersymmetry was constructed by Wess and Zumino \cite{Wess_Zumino}. It may play a fundamental role because the mathematical structure of supersymmetry called graded Lie superalgebra is the only one consistent with that in relativistic quantum field theory \cite{Haag1975}. The first realistic supersymmetric version of the standard model known as the minimal supersymmetric standard model was also later proposed to solve the hierarchy problem about the large discrepancy between aspects of the weak force and gravity \cite{Fayet1977}. Regardless of the intense theoretical works on supersymmetry in the particle physics, none of the superpartners, which have identical properties except for opposite statistics, of any known elementary particle has been found in experiments so far. Even though the Large Hadron Collider (LHC) recently confirmed the discovery of the Higgs particle \cite{Higgs_LHC, Englert, Higgs}, there is still no evidence of any supersymmetric particle.

Despite the difficulties in confirming the \textit{relativistic} SUSY in high energy physics, there have been a few theoretical proposals on the spacetime SUSY in condensed matter and atomic systems, e.g. the possible emergence of spacetime SUSY at quantum critical points in lattice models \cite{SSLee2007, Yao-SUSY}, simulation of the spacetime SUSY in optical lattices in atomic systems \cite{Yu2010}, and the emergence of spacetime SUSY at the boundary of topological phases in condensed matter systems\cite{Grover-science}. On a different front, \textit{nonrelativistic} SUSY (a Bose-Fermi symmetry unrelated to space-time symmetry) has attracted considerable interest due to the recent experimental progress in mixtures of ultra cold Bose and Fermi atoms \cite{Schreck2001,Hadzibabic2002,Roati2002,Modugno2003,Schori2004,Inouye2004,Goldwin2004,Stan2004}. The Bose-Fermi mixtures provide an opportunity to realize and study supersymmetry. An ultra cold superstring model was constructed \cite{Snoek2005,Snoek2006,Lozano2007}. The physical behavior of an exactly soluble of one-dimensional Bose-Fermi mixtures was investigated \cite{Imambekov2006,Imambekov_review,Lai1971}. A general formalism to study such supersymmetric systems based on coherent state path integral was also set up \cite{Mieck2007}. 

In Ref.~[\onlinecite{Yu2008}], one of the present authors studied general properties of supersymmetric Bose-Fermi mixtures, in which bosons and fermions are supersymmetric partners of each other. In the presence of time-reversal symmetry, supersymmetry is broken by a chemical potential difference between bosons and fermions, $\mu_f - \mu_b \equiv \Delta \mu >0$, which still keeps the canonical Hamiltonian, consisting of kinetic energy and the potential terms, supersymmetric. Such systems support a sharp fermionic collective excitation \cite{Yu2008} similar to that in supersymmetric high-energy theories, referred to as the Goldstino. The spectral function of the Goldstino mode has a sharp $\delta$-function peak at zero momentum ${\bf k}=0$ and frequency $\omega =\Delta \mu$. However, in experiments, it is a tremendous task to tune the parameters in such Bose-Fermi mixtures to keep the canonical Hamiltonian supersymmetric. The supersymmetry of the canonical Hamiltonian is expected to be broken and the sharp $\delta$-function peak at $\omega =\Delta \mu$ of the spectral function will be modified due to the {\it relaxation} of the Goldstino mode, which is the main focus in this work. 

Studying the relaxation of the Goldstino, a.k.a. the broadening of the sharp $\delta$-function peak at $\omega = \Delta \mu$ of the spectral function corresponding to the Goldstino mode, is not as straightforward as it seems, because the Goldstino mode itself is a highly nontrivial zero-momentum mode involving the linear combination of the fermion-boson pair at all sites (or all momentum) in the system. The present problem we encounter \textit{mimics} the electron spin resonance (ESR) with anisotropy in magnetic systems. Comparing the two problems, we are interested in the relaxation of the $q=0$ Goldstino mode, $Q\sim \sum_j f_j^\dagger b_j$, while the ESR measures the $q=0$ mode, e.g. motion of total spin in Heisenberg anti-ferromagnets (AFM), $S^+ \equiv \sum_j S_j^+$. We adopt the Kubo-Tomita (KT) theory \cite{Kubo-Tomita}, the well-known theoretical approach for the ESR, to perform a general analysis of the relaxation of the Goldstino mode. 

The paper is organized as follows. In Sec.~\ref{Sec:model} we introduce the model system followed by performing the general analysis of the broadening effects due to supersymmetry breaking using KT theory. We conclude the work in Sec.~\ref{Sec:conclusion}. We briefly review the KT theory in App. ~\ref{APP:KT_theory} and provide a perturbative analysis in the non-interacting limit in App.~\ref{APP:perturbation_U=0} which is consistent with the results in the more general KT theory approach.

\section{Kubo-Tomita theory approach to the relaxation of Goldstino mode}\label{Sec:model}
The model Hamiltonian is defined as $\hat{H} = \hat{H}_0 + \hat{H}' = \hat{H}_1 + \hat{H}_2 + \hat{H}'$, where
\begin{eqnarray}
&& \hat{H}_1 = - \mu_f \hat{N}_f - \mu_b \hat{N}_b, \\
\nonumber && \hat{H}_2 = \hat{T}+\hat{V} =\\
\nonumber &&= - t_h \sum_{\la j k \ra} \left[ b^\dagger_j b_k + f^\dagger_j f_k\right] + U \sum_j \left[ n^b_j n^f_j + \frac{1}{2} n^b_j \left( n_j^b -1 \right) \right],\\ \\
&&\hat{H}' =\delta U \sum_j \hat{n}_j^b \hat{n}_j^f.
\end{eqnarray}
The unperturbed Hamiltonian, $\hat{H}_0=\hat{H}_1 + \hat{H}_2$, contains the chemical potential term, $\hat{H}_1$, and the canonical Hamiltonian,$\hat{H}_2$, at the SUSY point. $\hat{H}'$ represents the SUSY breaking term. In the Bose-Fermi mixtures, there are \textit{only} two possible SUSY breaking terms--(i):the interaction strength difference between the onsite boson-fermi interaction and the onsite boson-boson interaction and (ii):the difference between the hopping strengths of bosons and fermions. We consider the former case (i) first and the whole discussions below can also be applied to the latter case (ii), which will be discussed near the end of the paper. The Goldstino field is defined as, \cite{Yu2008}
\begin{equation}
Q \equiv \frac{1}{\sqrt{N}}\sum_j f_j^\dagger b_j,
\end{equation}
where $N=N_f+N_b$ is the total number of fermions and bosons. 

Before discussing how the supersymmetry breaking term affects the Goldstino spectral function lineshape, we first point out how the present model system mimics the ESR problem in the Heisenberg AFM with anisotropy. For a Heisenberg AFM in the Zeeman magnetic field with anisotropy in the Heisenberg term, which breaks the spin rotation symmetry, the Hamiltonian is $H_{AFM} = H_{AF} + H_z + H'$, with
\begin{eqnarray}
&& H_{AF} = J \sum_{\la j,\ell \ra} \vec{S}_j \cdot \vec{S}_\ell,\\
&& H_z = - B_z \sum_j S^z_j,\\
&& H'_{ans} = \sum_{\la j\ell \ra} \sum_{\alpha=x,y,z} \delta J^\alpha_{jk} S^\alpha_j S^\alpha_\ell.
\end{eqnarray}
ESR only probes the $q=0$ total spin mode, $S^+=\sum_j S^+_j$, and without anisotropy $H'_{ans}$, the equation of motion for the total spin can be solved exactly as
\begin{eqnarray}
&& \frac{dS^+}{dt} = i \left[ H_{AF} + H_z ,S^+ \right] = -i B_z S^+ \\
&&\Rightarrow S^+ (t) = S^+(0) e^{- i B_z t}.
\end{eqnarray}
Therefore, the spectral function corresponding to the retarded total spin Green's function $G_S (t) = -i \theta(t) \la [ S^+(t), S^-(0)] \ra$ will show a $\delta$-function peak located at $\omega = B_z$. However, the presence of the anisotropy in the Heisenberg term $H'_{ans}$ will have effects on broadening the linewidth and shifting the peak center. \cite{Kubo-Tomita}

Comparing the Hamiltonian in the Goldstino problem with that in the Heisenberg AFM, we can have almost a one-to-one correspondence between these two problems. The chemical potential term $H_1$ in the Bose-Fermi mixtures is similar to the Zeeman energy $H_z$ in the Heisenberg AFM, the canonical Hamiltonian $H_2$ consisting of kinetic and interacting terms is similar to the isotropic Heisenberg term $H_{AF}$, and the supersymmetry breaking term $H'$ is similar to the anisotropy term $H'_{ans}$ which significantly changes the spectral function line shape. 
   
Back to the Goldstino problem, we first notice that several commutators can be greatly simplified,
\begin{eqnarray}
&& \left[ \hat{H}_1, \hat{H}_2 \right] = 0,~~\left[\hat{H}',\hat{H}_1 \right] = 0, \\
&&  \left[ Q, \hat{H}_2 \right] = \left[ Q^\dagger, \hat{H}_2 \right]=0,\\
&& \left[ Q, H_1 \right] = (\mu_f -\mu_b)Q \equiv \Delta \mu Q,
\end{eqnarray}
while $[H',H_2]$ and $[Q,H']$ remain complex and are not commutable.  

In order to apply the KT theory, which is briefly reviewed in App.~\ref{APP:KT_theory} we introduce the interaction picture representation of $Q$ as
\begin{eqnarray}
\nonumber Q^{(0)}(t) = e^{iH_0t} Q e^{-i H_0 t} &=& e^{i H_2 t} e^{i H_1 t} Q e^{-i H_2 t} e^{-i H_1 t}\\
&=& Q  e^{ - i \Delta \mu t}.
\end{eqnarray}
According to the KT theory, we can identity the Fourier component $Q(\omega = \Delta \mu) = Q$. On the other hand, the interaction picture representation of $H'$ is
\begin{eqnarray}
\nonumber H'(t) &=& e^{i H_0 t} H' e^{-i H_0 t} = e^{i H_2 t} e^{i H_1 t} H' e^{-i H_1 t} e^{-i H_2 t} \\
&=& e^{i H_2 t} H' e^{-i H_2 t} = H'(\omega =0; t) \equiv H'(t),
\end{eqnarray}
where we identity $H'(\omega=0) = H'$. The Goldstino field at time $t$, $Q(t)$, can be expanded in powers of the perturbation as $Q (t)= \sum_{n=0}^{\infty} Q^{(n)}(t)$, where the general expression of $n$-th order of $Q$ is
\begin{eqnarray}
\nonumber && Q^{(n)}(t) \\
\nonumber &&\equiv \frac{1}{i^n} \int_0^t dt_1 \cdots \int_0^{t_{n-1}}dt_n \left[Q^{(0)}(t);H'(t_1),\cdots, H'(t_n)\right], \\ \label{Eq:KT-Q}
\end{eqnarray}
where we introduce $[A;B,C,\cdots, K] \equiv [\cdots[[A,B],C],\cdots, K]$. The retarded Goldstino Green's function can be expanded as 
\begin{eqnarray}
G^R (t) &\equiv& -i \theta(t) \left\la 0 \big{|} \left\{ Q(t), Q^\dagger(0) \right\} \big{|}0 \right\ra = \sum_{n=0}^\infty G^R_n(t),~~\label{Eq:KT-G}
\end{eqnarray}
where $\big{|}0\ra$ represents the (unperturbed) ground state. According to Eqs.~(\ref{Eq:KT-Q})-(\ref{Eq:KT-G}), we can extract the Goldstino Green's function up to any order and its corresponding spectral function. 

At zeroth order, we get 
\begin{eqnarray}
G^R_0(t) = -i \theta(t) e^{-i \Delta \mu t} \left\la \left\{ Q, Q^\dagger \right\} \right\ra =- i \theta(t) e^{-i \Delta \mu t}.~~
\end{eqnarray}
Introducing Fourier transform, $G^R_0(\omega) = \int dt G^R_0(t)e^{i\omega t}$, we get
\begin{eqnarray}
G^R_0(\omega) = \frac{1}{\omega - \Delta \mu + i 0^+},
\end{eqnarray}
and the spectral function at zero order is, $A^{(0)}(\omega) = -(1/\pi) Im G^R_0(\omega)$,
\begin{eqnarray}
A^{(0)}(\omega) = \delta (\omega - \Delta \mu),
\end{eqnarray}
which shows a $\delta$-peak located at $\Delta \mu$. The zeroth order (unperturbed) result is consistent with that in Ref.~[\onlinecite{Yu2008}].

For the first-order term, we have
\begin{eqnarray}
\nonumber G^R_1 (t)  &=& - i \theta(t) e^{-i \Delta \mu t} \frac{1}{i}\int_0^t dt' \left\la\left\{ \left[ Q,H'(t') \right],Q^\dagger \right\} \right\ra \\
\nonumber &=& - i \theta(t) e^{-i \Delta \mu t} \frac{1}{i}\int_0^t dt' \left\la\left\{ \left[ Q,H' \right],Q^\dagger \right\} \right\ra  \\
&=& - i \theta(t) e^{-i \Delta \mu t} (-it)  \left\la\left\{ \left[ Q,H' \right],Q^\dagger \right\} \right\ra, \end{eqnarray}
where from first line to second line we use the fact that $H_2$ commutes with $Q$. Using the identity $\{ Q, Q^\dagger\} =1$, we obtain that $\la\{ [Q,H'],Q^\dagger \}\ra$ is real, which leads to the fact that the effect of the first-order term is only to \textit{shift} the location of the $\delta$-function peak of the spectral function, which will be presented below. For a more concrete result, we crudely mean-field decouple the expectation value,
\begin{eqnarray}
\left\la \left\{ \left[Q,H'\right],Q^\dagger\right\}\right\ra \simeq -2\delta U \left\la n_j^b\right\ra \left\la \left\{ Q, Q^\dagger\right\}\right\ra = -2\delta U\rho_b,~~~~~~
\end{eqnarray}
where in the last equality we assume the density of the bosons is uniform and replace it by $\la n^b_j \ra = N_b/V \equiv \rho_b$, with $V$ being the whole system volume. After the mean-field decoupling, at first order we get the renormalized Goldstino Green's function
\begin{eqnarray}
\nonumber G^{R, (1)}(t) = G^R_0 (t) + G^R_1(t) &\simeq&  -i \theta (t) e^{-i \Delta \mu t} \left( 1 + i 2\delta U \rho_b t \right) \\
& \simeq & - i \theta (t) e^{-i \Delta \mu t} e^{+i 2\delta U \rho_b t},
\end{eqnarray}
where in the last line, we assume small $\delta U$ and exponentiate the correcting term. In Fourier space, we get
\begin{eqnarray}
G^{R,(1)}(\omega) \simeq \frac{1}{\omega -( \Delta \mu - 2 \rho_b \delta U)+ i 0^+},
\end{eqnarray}
which gives the spectral function
\begin{eqnarray}
A^{(1)}(\omega) \simeq \delta (\omega - ( \Delta\mu - 2\rho_b \delta U)).
\end{eqnarray}
The first-order perturbation indeed only shifts the $\delta$-function peak. The result is consistent with the previous studies.\cite{Shi2010} In App.~\ref{APP:perturbation_U=0}l we perform the perturbative studies in the $U=0$ limit and we also obtain exactly the same result at first-order.

For the second order term, the general form is
\begin{eqnarray}
\nonumber && G^R_2(t) \\
\nonumber && = i \theta(t) e^{-i \Delta \mu t} \int_0^t dt_1 \int_0^{t_1} dt_2 \left\la \left\{ \left[ Q;H'(t_1), H'(t_2)\right], Q^\dagger \right\}\right\ra \\
 && = i \theta(t) e^{-i \Delta\mu t}\int_0^t d\tau (t-\tau) \left\la \left\{ \left[ Q;H'(\tau), H' \right], Q^\dagger \right\}\right\ra . 
\end{eqnarray}
If we follow KT theory to approximate 
\begin{eqnarray}
\nonumber \left\la \left\{ \left[Q; H'(\tau),H'\right],Q^\dagger\right\}\right\ra &\simeq& \left\la \left\{ \left[Q,H'(\tau)\right],\left[ H',Q^\dagger\right] \right\}\right\ra \\
&\equiv& \sigma_G^2 f_G(\tau),
\end{eqnarray}
with
\begin{eqnarray}
&& \sigma_G^2 \equiv \left\la \left\{ \left[ Q, H'\right],\left[ H', Q^\dagger \right] \right\} \right\ra,\\
&& f_G(\tau) \equiv \frac{ \left\la \left\{ \left[Q,H'(\tau)\right],\left[ H',Q^\dagger\right] \right\}\right\ra}{ \left\la \left\{ \left[Q,H'\right],\left[ H',Q^\dagger\right] \right\}\right\ra},
\end{eqnarray}
we can approximately rewrite $G^R_2(t)$ as
\begin{eqnarray}
G^R_2(t)\simeq -G^R_0(t) \sigma_G^2\int_0^t d\tau (t-\tau) f_{G}(\tau).
\end{eqnarray}
It is clear that $\sigma_G^2 \propto (\delta U)^2$ and $f(\tau)$ is a ``dimensionless'' function which depends on variable $\tau$, and $\sigma_G^2$ and $f_G(\tau)$ are both positive and real in this approximation. For the ``time-independent'' $\sigma_G^2$, we again crudely mean-field decouple it as
\begin{eqnarray}
\sigma_G^2=\left\la \left\{ \left[ Q, H'\right],\left[ H',Q^\dagger \right]\right\} \right\ra \simeq\left(\delta U\right)^2 \left( 6 \rho_b^2 + 2 \rho_b\right).
\end{eqnarray}

If we know the exact or approximate form of $f_G(\tau)$, we can get the exact or approximate form of $G^R_2(t)$. We do not find a way to get the exact form of $f_G(\tau)$, and below we take a plausible form of $f_G(\tau)$ \cite{Bloembergen1948}, which was proposed in the studies in relaxation effects in nuclear magnetic resonance absorption, that leads to a reasonable result.

Before discussing the general analysis of the second order term, we first point out that the main difficulty of extracting the exact form of $f_G(\tau)$ is that the time dependence of $H'(\tau)$ can not be factored out similar to that for $Q^{(0)}(\tau)$ due to the complicated form of $[H', H_2]$. Nevertheless, for a \textit{qualitative} result in the ``long-time'' limit, $t\rightarrow \infty$, we can approximate 
\begin{eqnarray}
 G^R_2(t) &\simeq& -G^R_0(t) \sigma_G^2 t \int_0^{\infty} d\tau f_G(\tau)\equiv  -G^R_0(t) \sigma_G^2 t \tau',~~
\end{eqnarray}
where $\tau'\equiv \int_0^\infty d\tau f_G(\tau)$. The renormalized Goldstino Green's function containing up to second-order perturbation is
\begin{eqnarray}
\nonumber G^{R,(2)}(t) &=& G_0(t) + G_1(t) + G_2(t)\\
\nonumber &\simeq& G^R_0(t)\left(1+ i 2\rho_b \delta U -\sigma_G^2 \tau' t\right)\\
&\simeq& G^R_0(t) e^{i(2\rho_b \delta U)t- \sigma_G^2\tau' t},
\end{eqnarray}
which gives the spectral function 
\begin{eqnarray}
A^{(2)}(\omega) \simeq \frac{1}{\pi} \frac{\sigma_G^2\tau'}{\left[\omega -(\Delta\mu - 2 \rho_b \delta U)\right]^2 + (\sigma_G^2\tau')^2}.
\end{eqnarray}
We can see that the lineshape is a Lorentzian function with width $\sigma_G^2\tau' \propto (\delta U)^2$ centered around $\Delta\mu- 2\rho_b \delta U$. $\sigma_G^2 \tau'$ corresponds to the ``relaxation-rate'' of the Goldstino or we can define a Goldstino ``relaxation time'' via the relation $1/T_G = \sigma_G^2 \tau' \propto (\delta U)^2$. For further strengthening the conjecture of the Lorentzian lineshape of the spectral function of Goldstino Green's function, in App.~\ref{APP:perturbation_U=0}l we provide the perturbative calculations up to second-order of $\delta U$ in the limit $U=0$. We indeed find that the second-order self-energy contains an imaginary part proportional to square of the perturbation, $Im[\Sigma^{(2)}] \propto (\delta U)^2$, and the lineshape of the spectral function of Goldstino Green's function is modified to be Lorentzian with width $\sim Im[\Sigma^{(2)}]\propto (\delta U)^2$, consistent with the results presented here using KT theory approach.

In the general circumstance, it is difficult to get an analytical result unless we know the exact form of $f_G(\tau)$. Since $f_G(\tau)$ is a dimensionless function, the simplest form it can be, which is well-defined at any time, is \cite{Bloembergen1948}
\begin{eqnarray}
f_G(\tau) = e^{- \tau/\tau_0},
\end{eqnarray}
where $\tau_0$ is some ``characteristic time" with unit of energy inverse (the same unit to time since we set $\hbar \equiv 1$). The simplest form of $\tau_0$ is a function consisting of the interaction strength inverse $U^{-1}$, the hopping strength inverse $t_h^{-1}$, and the chemical potentials of fermions and bosons, $\mu_{f/b}$, $\tau_0 = h(U^{-1}, t_h^{-1},\mu_f^{-1},\mu_b^{-1})$. With the form of $f_G(\tau)$, we get
\begin{eqnarray}
G^R_2(t) &\simeq -G^R_0(t) \sigma_G^2 \tau_0^2 \left[ e^{-\frac{t}{\tau_0}}-1+ \frac{t}{\tau_0} \right]. \label{Eq:2nd-order}
\end{eqnarray}
Therefore,
\begin{eqnarray}
\nonumber && G^{R,(2)}(t) \simeq  G^R_0(t) e^{i 2\rho_b \delta U}e^{- \sigma_G^2 \tau_0^2 \left( e^{-t/\tau_0} -1 + \frac{t}{\tau_0}\right)}\\
\nonumber &&\simeq e^{\sigma_G^2\tau_0^2}G^R_0(t) e^{i 2 \rho_b \delta U}e^{-\sigma_G^2 \tau_0 t} \sum_{n=0}^\infty \frac{1}{n!}\left( - \sigma_G^2 \tau_0^2 e^{-t/\tau_0}\right)^n.  \\
\end{eqnarray}
The series converge very rapidly for $\sigma_G \tau_0 \le 1$. The leading term with $n=0$ gives the spectral function 
\begin{eqnarray}
A^{(2)}(\omega) \simeq \frac{e^{\sigma_G^2 \tau_0^2}}{\pi}\frac{\sigma_G^2 \tau_0}{\left[\omega-\left(\Delta \mu - 2\rho_b \delta U\right)\right]^2 + (\sigma_G^2 \tau_0)^2},~~~
\end{eqnarray}
which is a Lorentzian function with half-width $\sigma_G^2 \tau_0 \sim (\delta U)^2 h(U^{-1}, t_h^{-1},\mu_f^{-1},\mu_b^{-1}) \propto (\delta U)^2$.

If $\sigma_G\tau_0 \rightarrow \infty$, which means $\tau_0 \rightarrow \infty$ since $\sigma_G \sim \rho_b \delta U <1$, Eq.~(\ref{Eq:2nd-order}) can be Taylor expanded in powers of $t/\tau_0$ giving the leading term,
\begin{eqnarray}
G_2^R(t) \overrightarrow{\sigma_G\tau_0 \rightarrow \infty}  - G^R_0(t) \sigma_G^2 \frac{t^2}{2}.
\end{eqnarray}
Note that in this limit, the leading second-order term is $\tau$-independent. If we ignore higher-order terms in the Taylor series, the leading result corresponds to the approximation that $[H', H_2]\simeq 0$, which means the time-dependence of $H'(\tau)$ is dropped. In this limit, the renormalized Goldstino Green's function becomes
\begin{eqnarray}
G^R_2(t) \overrightarrow{\sigma_G \tau_0 \rightarrow \infty} G^R_0(t) e^{-i 2\rho_b \delta U -\sigma_G^2 t^2/2},
\end{eqnarray}
which gives spectral function 
\begin{eqnarray}
A^{(2)}(\omega) \overrightarrow{\sigma_G \tau_0 \rightarrow \infty} \frac{1}{\sqrt{2\pi \sigma_G^2}} e^{-\frac{[\omega - (\Delta \mu - 2 \rho_b \delta U)]^2}{2 \sigma_G^2}}.
\end{eqnarray}
The spectral function lineshape becomes a ``Gaussian distribution'' with width (standard deviation) $\sigma_G\propto |\delta U|$, which is consistent with the first situation originally discussed by Kubo and Tomita and briefly reviewed in the App.~\ref{APP:KT_theory}, in which the Fourier components of perturbation, i.e. $H'(\omega)$, are time-independent.

The crossover of the lineshape from a Lorentzian function to a Gaussian distribution can be possibly observed experimentally. In the present case, the commutator, $[H', H_2] = [H', \hat{T}]\not=0$, does not vanishes because the perturbation, $H'$, is not commutable with the hopping term, $\hat{T}$. In the cold atom systems, it is possible to gradually decrease the hopping strength by tuning the potential depth \cite{Jaksch1998, Duan2003, Bloch2008}, which makes it possible to observe experimentally the change of the lineshape and the crossover from a Lorentzian with half-width $\propto (\delta U)^2$ to a Gaussian distribution with width $\propto |\delta U|$ before entering the Mott-insulating phase.

So far we have focused on the case in which the SUSY breaking term is the difference between the strength of the onsite fermion-boson and onsite boson-boson interaction. On the other hand, if we consider the other possible SUSY breaking term, case (ii):the hopping strength difference between bosons and fermions, whose Hamiltonian is
\begin{eqnarray}
H'_h =\delta t_h \sum_{\la j k\ra} f_j^\dagger f_k,
\end{eqnarray}
most of the discussions above can be directly applied to this case with $H'\rightarrow H'_h$. For the first-order term, we find that 
\begin{eqnarray}
\left\la \left\{ \left[ Q, H'_h \right], Q^\dagger \right\}\right\ra= \frac{\delta t_h}{N}\sum_{\bf k} \xi_{\bf k} \left( \big{\la} n_{\bf k}^b \big{\ra} + \big{\la} n_{\bf k}^f \big{\ra} \right) \equiv  \delta t_h \gamma,~~~~~~
\end{eqnarray}
where we introduce $\xi_{\bf k} \equiv -\sum_{\{ {\bf e}_\mu \}} e^{-i {\bf k}\cdot {\bf e}_\mu}$, with $\{{\bf e}_\mu\}$ being the unit vectors that connect a site to its nearest neighbor sites. Following the previous discussions, we obtain, at the first order, the spectral function as $
A^{(1)}(\omega) \simeq \delta\left( \omega - \left( \Delta \mu + \gamma \delta t_h\right)\right)$, which shows the shift of the $\delta$-peak. The second-order term involves 
\begin{eqnarray}
\nonumber \left\la \left\{ \left[ Q, H'_h\right],\left[ H'_h, Q^\dagger \right] \right\}\right\ra &=& \frac{ (\delta t_h)^2}{N}\sum_{\bf k} \xi'_{\bf k} \left( \big{\la} n^b_{\bf k}\big{\ra} + \big{\la} n_{\bf k}^f \big{\ra}\right)\\
&\equiv& (\delta t_h)^2 \gamma',~~~~
\end{eqnarray}
where we introduce $\xi'_{\bf k} \equiv \sum_{\bf k} \sum_{\{ {\bf e}_\mu \}}\sum_{\{ {\bf e}_\nu \}} e^{i {\bf k}\cdot ({\bf e}_\mu - {\bf e}_\nu)}$. Following the previous discussions, we conclude that $\sigma_G^2 \sim (\delta t_h)^2$ and the spectral function becomes a Lorentzian function with half-width $\sigma_G^2 \tau_0 \sim (\delta t_h)^2 h'(U^{-1}, t_h^{-1},\mu_f^{-1},\mu_b^{-1})$, where $h'(x,y, z,w)$ is a phenomenology function that has the correct unit. Furthermore, if we experimentally ``tune down'' the interaction strength to make $[ H'_h, H_2] = [H'_h, \hat{V}] \rightarrow 0$, the time-dependence of $H'_h(t)$ can be dropped. The spectral function lineshape in this limit becomes a Gaussian distribtution with width $\propto |\delta t_h|$, which can be experimentally confirmed. 
\section{Conclusion}\label{Sec:conclusion}
We study the relaxation of Goldstino mode in Bose-Fermi mixtures due to the supersymmetry breaking. We adopt the well-known Kubo-Tomita theory in electron spin resonance theory to find that the spectral function of Goldstino Green's function, in general, is broadened to be a Lorentzian function with width proportional to the perturbation strength square. In the limit where either the hopping strength or the interaction strength vanishes, the lineshape becomes a Gaussian distribution with width linearly proportional to the perturbation strength.

\acknowledgments
This research is supported by the National Science Foundation through grants No. DMR-1004545 and No. DMR-1442366.
\appendix

\section{Brief Review of Kubo-Tomita Theory}\label{APP:KT_theory}
Most of the physical quantities can be obtained from the corresponding Green's functions $G(t)$, e.g. retarded Green's function, real-time ordered Green's function, imaginary-time Green's function depending on the convention you use. The typical way to examine the resonance ``frequency'' is to examine the corresponding Green's function in Fourier space, $G(\omega) = \int d\omega G(t) e^{i\omega t}$. In the noninteracting limit, $H = H_0$, the Fourier components give sharp resonance peaks at certain frequencies, $G(\omega) = \sum_\alpha G(\omega_\alpha)\delta(\omega- \omega_\alpha)$, and this means the form of the Green's function at time domain should be $G(t) = \sum_\alpha G_\alpha e^{- i \omega_\alpha t}$. In the presence of (weak) interaction $H'$, the sharp resonance lines will get shifted and broadened. The conventional way to examine the effects of the interaction is to use equation of motion theory to expand the Green's function in powers of $t$. Defining the retarded Green's function $G(t) = -i \theta(t) \la O(t) O^\dagger (0)\ra_0$, we can use equation of motion theory to expand $O(t)$ as
\begin{eqnarray}
\nonumber O(t) = O(0) + \frac{t}{i} [ O(0),H] + \cdots + \frac{t^n}{n!}[O(0); H,\cdots, H], \\
\end{eqnarray} 
where we introduce $[A; B, C,\cdots ,K] \equiv [\cdots[[ A,B],C]\cdots ,K]$. The results above can be obtianed as follows. From Heisenberg equation of motion, we know
\begin{eqnarray}
\frac{d O(t)}{dt} = \frac{1}{i} [ O(t),H].
\end{eqnarray}
Integrating both sides with the initial condition $O(t=0)=O(0)$, we obtain
\begin{eqnarray}
O(t) = O(0) + \int_0^t dt_1 [ O(t_1),H].
\end{eqnarray}
By iteration, we get the result above. We then obtain
\begin{eqnarray}
\nonumber G(t) &=& -i \theta(t)  \sum_n \frac{t^n}{n!} \bigg{\la}\bigg{\{} \big{[}O(0);H,\cdots,H\big{]}, O^\dagger(0)\bigg{\}}\bigg{\ra}\\
&=&  G_0(t) +  G_1(t) + G_2(t) + \cdots + G_n(t).
\end{eqnarray}
Despite the simplicity of the expression above, it is, in most cases, difficult to analyze it analytically. The main reason is that we are interested in extracting the results perturbatively in powers of $H'$, but, however, it is not easy to extract the perturbative results in powers of $H'$ in this approach. Besides, if we go to the noninteracting limit, where the strength of the perturbation $H' \propto \epsilon$ vanishes, the perturbative result should go back to the noninteracting result as
\begin{eqnarray}
lim_{\epsilon \rightarrow 0} G(t) = \sum_\alpha G_\alpha e^{-i \omega_\alpha t}.
\end{eqnarray}
In the presence of perturbation $H'$, we may assume generally that the coefficients of the exponential functions become function of $t$,
\begin{eqnarray}
G(\epsilon, t) = \sum_\alpha G_\alpha (\epsilon ,t) e^{-i \omega_\alpha t},
\end{eqnarray}
which gives the limits
\begin{eqnarray}
lim_{\epsilon \rightarrow 0} G_\alpha (\epsilon, t) = G_\alpha,
\end{eqnarray}
which is independent of $t$.

For extracting the results perturbatively in powers of $H'$, Kubo and Tomita (KT) introduced another way of performing the expansion in powers of the perturbation $H'$. Below, we briefly review the KT theory.\\

As illustrated above, the most convenient way to obtain the expansion of $G(t)$ is to solve the equation of motion,
\begin{eqnarray}
i \frac{d O(t)}{dt} = \left[ O(t) ,H\right] = \left[ O(t) , H_0 + H'\right].
\end{eqnarray}
We are interested in performing the expansion in powers of $H'$, and in order to achieve that, we switch to the interaction picture and introduce
\begin{eqnarray}
O(t) = e^{i H_0 t} O^*(t) e^{-i H_0 t}, \label{Eq:int_pic}
\end{eqnarray}
where $H_0$ is the unperturbed Hamiltonian. In the interaction picture, we obtain the equation of motion,
\begin{eqnarray}
i\frac{d O^*(t)}{dt} = \left[ O^*(t),H'(-t) \right],
\end{eqnarray}
where we also introduce $H'(t) = e^{i H_0 t} H' e^{-i H_0 t}$. Integrating both sides of the equation with the initial condition $O^*(t=0) = O(0)\equiv O$ and we get
\begin{eqnarray}
O^*(t) =O + \frac{1}{i} \int_0^t dt_1 \left[ O^*(t_1),H'(-t_1)\right].
\end{eqnarray}
By iteration, we get
\begin{eqnarray}
\nonumber O^*(t) &=& O + \sum_{n=1}^\infty (i)^{-n} \int_0^t dt_n \int_0^{t_n} dt_{n-1} \cdots \int_0^{t_2} dt_1 \times \\
&& \hspace{2cm} \times \left[ O; H'(-t_1),\cdots, H'(-t_n)\right].
\end{eqnarray}
Plugging the expansion above to Eq.~(\ref{Eq:int_pic}) gives
\begin{eqnarray}
O(t) = O^{(0)}(t) + O^{(1)}(t) + \cdots + O^{(n)}(t)+ \cdots,
\end{eqnarray}
where the general expression of $O^{(n)}(t)$ is
\begin{eqnarray}
\nonumber O^{(n)}(t) &=& (i)^{-n} \int_0^t dt_1 \int_0^{t_1} dt_2 \cdots \int_0^{t_{n-1}}dt_n \times \\
&& \hspace{1cm} \times \left[ O^{(0)}(t); H'(t_1),\cdots,H'(t_n)\right],
\end{eqnarray}
with
\begin{eqnarray}
&& O^{(0}(t) =e^{i H_0 t} O e^{-iH_0t},\\
&& H'(t) = e^{i H_0 t} H' e^{-i H_0 t}.
\end{eqnarray}
With the expansion of $O(t)$ in the powers of $H'$, the Green's function can be straightforwardly expanded as
\begin{eqnarray}
G(t) = G_0(t) + G_1(t) + G_2(t) + \cdots,
\end{eqnarray}
where
\begin{eqnarray}
\nonumber G_n(t) &=& (i)^{-n} \int_0^t dt_1 \int_0^{t_1} dt_2 \cdots \int_0^{t_{n-1}} dt_n (-i)\theta(t)  \times\\
\nonumber && \times \bigg{\la} \bigg{\{} [O^{(0)}(t); H'(t_1),\cdots, H'(t_n)],O^\dagger(0) \bigg{\}}\bigg{\ra}. \\
\end{eqnarray} 
The result we obtain above is from straightforward expansion in orders of the perturbation $H'$ without making many assumptions, except for  assuming $H'$ can be treated as a perturbation. Hence, the general formula above should be valid in most (if not all) of the situations.\\

The general way of evaluating the expression is to expand $O^{(0)}(t)$ and $H'(t)$ in a Fourier series. We note that it is not always possible to decompose $O^{(0)}(t)$ and $H'(t)$ into Fourier series exactly, but, fortunately, in most problems it is possible. There are two general situations depending on whether or not the Fourier components of $O^{(0)}(t)$ and $H'(t)$, i.e., $O(\omega_\alpha)$ and $H'(\omega_\alpha)$, develop time dependence. Let us first focus on the first situation in which the Fourier components are ``time-independent'', and the generalization to the second case can be straightforward carried over once we know all the logics developing in the first case.\\
 
In the first case, the $O^{(0)}(t)$ and $H'(t)$ can be fully decomposed to a Fourier series,
\begin{eqnarray}
&& O^{(0)}(t) = \sum_\alpha O(\omega_\alpha) e^{-i \omega_\alpha t},\\
&& H'(t) = \sum_\alpha H'(\omega_\alpha) e^{-i \omega_\alpha t},
\end{eqnarray}
where the Fourier components $O(\omega_\alpha)$ and $H'(\omega_\alpha)$ are  ``time independent''. We can use the identity
\begin{eqnarray}
\nonumber e^A B e^{-A} &=& B + [A,B]+\frac{1}{2!}[A,[A,B]]+\frac{1}{3!}[A,[A,[A,B]]]\\
&&+\cdots
\end{eqnarray}
to get
\begin{eqnarray}
&& [O(\omega_\alpha),H_0] = \omega_\alpha O(\omega_\alpha),\\
&& [H'(\omega_\alpha,H_0]=\omega_\alpha H'(\omega_\alpha).
\end{eqnarray}
Then we can repress $G_n(t)$ as
\begin{widetext}
\begin{eqnarray}
G_n(t) = -i\theta(t) \sum_\alpha i^{-n} e^{-i \omega_\alpha t} \sum_{\omega_1}\cdots \sum_{\omega_n} g_n(t;\omega_1,\cdots,\omega_n)F_n(\omega_\alpha;\omega_1,\cdots,\omega_n),
\end{eqnarray}
where
\begin{eqnarray}
&& g_n(t;\omega_1,\cdots,\omega_n) \equiv \frac{1}{2\pi i} \int_{-i\infty}^{i\infty} e^{pt} [ p (p+i\omega_1)(p+i\omega_1 + i \omega_2) \cdots (p + i \sum_{\ell =1}^{n} \omega_\ell)]^{-1}dp,\\
&& F_n(\omega_\alpha;\omega_1,\cdots,\omega_n) \equiv \bigg{\la} \bigg{\{} \left[ O(\omega_\alpha);H'(\omega_1), \cdots ,H'(\omega_n)\right],O^\dagger(0)\bigg{\}}\bigg{\ra}.
\end{eqnarray}
\end{widetext}
We also know that $O^\dagger(t=0) = \sum_\beta O^\dagger(\omega_b)$, and we can introduce
\begin{eqnarray}
\nonumber &&F_n(\omega_\alpha; \omega_1,\cdots, \omega_n) \\
\nonumber &&= \sum_\beta \bigg{\la} \bigg{\{} \left[ O(\omega_\alpha);H'(\omega_1), \cdots, H'(\omega_n)\right],O^\dagger(\omega_\beta)\bigg{\}}\bigg{\ra} \\
&&\equiv \sum_\beta F_n^*(\omega_\alpha,\omega_\beta;\omega_1,\cdots, \omega_n).
\end{eqnarray}
From conservation of frequency, we know $F_n^*$ vanishes unless
\begin{equation}
\omega_\alpha-\omega_\beta + \sum_{\ell=1}^n \omega_\ell =0.
\end{equation}
The reason that we introduce the function $g_n(t;\omega_1,\cdots,\omega_n)$ is due to the fact the it is easier to analyze the $g_n$ function, whose result can be obtained by summing the residues of the simple poles located at $p=0,~-i\omega_1,~-i\omega_1-i\omega_2, \cdots,$ and etc. We can then write down the general expression along with the constraint of $G_n(t)$,
\begin{eqnarray}
&&\nonumber G_n(t) =-i\theta(t) \sum_{\alpha,\beta} i^{-n} e^{-i \omega_\alpha t} \sum_\omega g_n(t;\omega)F^*_n(\omega_\alpha,\omega_\beta;\omega)\\
&& \hspace{3.5 cm} \times \delta_{\omega_\alpha - \omega_\beta + \sum_{\ell=1}^n\omega_\ell=0},
\end{eqnarray}
where we introduce the abbreviations, $\sum_\omega \equiv \sum_{ \omega_1}\cdots \sum_{\omega_n}$, $g_n(t;\omega) \equiv g_n(t;\omega_1,\cdots, \omega_n),$ and $F^*(\omega_\alpha, \omega_\beta; \omega) \equiv F^*(\omega_\alpha, \omega_\beta; \omega_1,\cdots, \omega_n)$. Let's try to examine some simple cases. For the first order perturbation, we have
\begin{eqnarray}
\nonumber && G_1(t) = -i\theta(t) \sum_{\alpha,\beta} e^{-i\omega_\alpha t} \sum_{\omega_1} \frac{e^{-i\omega_1 t}-1}{\omega_1} \times\\ 
\nonumber && \hspace{3 cm} \times F^*_1(\omega_\alpha,\omega_\beta;\omega_1) \delta_{\omega_\alpha -\omega_\beta + \omega_1 =0} \\
\nonumber &&= -i\theta(t) \sum_\alpha e^{-i \omega_\alpha t} \bigg{[} -i t F_1^*(\omega_\alpha, \omega_\alpha;0) + \\
\nonumber &&\hspace{0.5 cm}+ \sum_\beta \frac{1 - e^{i (\omega_\alpha - \omega_\beta )t}}{\omega_\alpha - \omega_\beta} F_1^*(\omega_\alpha,\omega_\beta;-(\omega_\alpha - \omega_\beta))\bigg{]}\\
\nonumber && = -i\theta(t) \sum_\alpha e^{-i \omega_\alpha t} \bigg{[} -i t F_1^*(\omega_\alpha, \omega_\alpha;0) + \\
\nonumber && \hspace{1cm} +  \sum_\beta \frac{1}{\omega_\alpha - \omega_\beta} \big{(} F_1^*(\omega_\alpha,\omega_\beta;-(\omega_\alpha - \omega_\beta))+ \\
&&\hspace{3 cm}+ F_1^*(\omega_\beta,\omega_\alpha;\omega_\alpha - \omega_\beta)\big{)}\bigg{]},
\end{eqnarray}
where 
\begin{eqnarray}
&& F_1^*(\omega_\alpha,\omega_\alpha;0) = \bigg{\la}\bigg{\{} \big{[} O(\omega_\alpha),H'(0)\big{]},O^\dagger(\omega_\alpha)\bigg{\}}\bigg{\ra},\\
\nonumber &&F_1^*(\omega_\alpha,\omega_\beta;-(\omega_\alpha-\omega_\beta)) + F_1^*(\omega_\beta;\omega_\alpha;\omega_\alpha-\omega_\beta)\\
\nonumber && = \bigg{\la}\bigg{\{} \big{[} O(\omega_\alpha),H'(\omega_\beta-\omega_\alpha)\big{]},O^\dagger(\omega_\beta)\bigg{\}}+\\
&& \hspace{1cm} + \bigg{\{}\big{[} O(\omega_\beta),H'(\omega_\alpha -\omega_\beta)\big{]},O^\dagger(\omega_\alpha)\bigg{\}}\bigg{\ra}.
\end{eqnarray}
Using the fact $H'^\dagger (\omega) =  H'(-\omega)$ and similarly $O^\dagger(\omega) = O(-\omega)$, we see that the terms in $[...]$ are all real, which simply indicates that the first order perturbation will only give resonance line ``shift'' in the Fourier space. The second-order perturbation contains more complex terms. The general expression is
\begin{eqnarray}
\nonumber G_2(t)&=& -i\theta(t) \sum_{\alpha,\beta}\frac{e^{-i\omega_\alpha t}}{i^2}\sum_\omega g_2(t;\omega_\gamma,\omega_\delta) \times \\
&& \hspace{0.1 cm} \times F^*_2(\omega_\alpha,\omega_\beta;\omega_\gamma,\omega_\delta) \times \delta_{\omega_\alpha-\omega_\beta + \omega_\gamma + \omega_\delta=0}.
\end{eqnarray}
There are several choices for satisfying the condition. $(1):~\omega_\alpha=\omega_\beta,~\omega_\gamma =0=\omega_\delta,$ $(2):~\omega_\alpha=\omega_\beta,~\omega_\gamma = -\omega_\delta \not=0.$ $(3): \omega_\alpha \not =\omega_\beta,~\omega_\delta=0, \omega_\gamma = -(\omega_\alpha-\omega_\beta).$ $(4):~\omega_\alpha\not=\omega_\beta,~\omega_\gamma=0, \omega_\delta = -(\omega_\alpha - \omega_\beta).$ $(5): \omega_\delta = - (\omega_\alpha -\omega_\beta + \omega_\gamma) \not=0$. Below, we will only keep the terms with $ \omega_\alpha = \omega_\beta$ and $\omega_\gamma = -\omega_\delta$ and ignore other highly asymmetric terms with $\omega_\alpha \not=\omega_\beta$. The result ignoring highly asymmetric terms is
\begin{eqnarray}
\nonumber G_2(t) \simeq && -i\theta(t) \sum_\alpha e^{-i \omega_\alpha t} \bigg{[} - \frac{t^2}{2}F_2^*(\omega_\alpha, \omega_\alpha;0,0) - \\
\nonumber && -\sum_{\gamma\not=0} \frac{1- i \omega_\gamma t - e^{-i \omega_\gamma t}}{\omega_\gamma^2}F_2^*(\omega_\alpha, \omega_\alpha; \omega_\gamma, - \omega_\gamma)\bigg{]}. \\
\end{eqnarray}
Since we already know that the first-order perturbation only shift the location of the resonance line, let us see focus on the effects of second-order perturbation due to $G_2(t)$. Combining the zeroth order term with the second order term, we obtain
\begin{eqnarray}
\nonumber && G^{(2)}(t) \simeq G_0(t) + G_2(t) \\
\nonumber && \simeq - i \theta(t) \sum_\alpha e^{-i \omega_\alpha t} F^*_0(\omega_\alpha, \omega_\alpha) \bigg{[} 1- \sum_{\gamma \not=0} \frac{1 - i \omega_\gamma t}{\omega_\gamma^2} \sigma_{\alpha \gamma}^2 - \frac{t^2}{2}\sigma_{\alpha 0}^2\bigg{]}-\\
\nonumber &&\hspace{0.5cm}- i \theta(t) \sum_{\alpha}\sum_{\gamma\not=0} \frac{F_0^*(\omega_\alpha,\omega_\alpha)}{\omega_\gamma^2} e^{-i (\omega_\alpha + \omega_\gamma)t}\sigma_{\alpha \gamma}^2 \\
\nonumber && \simeq -i\theta(t) \sum_\alpha  F_0^*(\omega_\alpha, \omega_\alpha) \bigg{(} 1- \sum_{\gamma \not=0} \frac{\sigma_{\alpha \gamma}^2}{\omega_\gamma^2}\bigg{)}e^{-i (\omega_\alpha-\sum_{\gamma\not=0}\frac{\sigma_{\alpha \gamma}^2}{\omega_\gamma} )t- \frac{\sigma_{\alpha 0}^2}{2}t^2} \\
&& \hspace{0.5cm} - i \theta(t) \sum_{\alpha, \gamma\not=0} \frac{\sigma_{\alpha \gamma}^2F_0^*(\omega_\alpha, \omega_\alpha)}{\omega_\gamma^2} e^{-i(\omega_\alpha + \omega_\gamma)t}.
\end{eqnarray}
where we define $\sigma_{\alpha \gamma}^2 \equiv F_2^*(\omega_\alpha, \omega_\alpha; \omega_\gamma, - \omega_\gamma)/F_0^*(\omega_\alpha,\omega_\alpha)$. The matrix element in the frequency space forms a trace and we can use the trace identity to rewrite
\begin{eqnarray}
\sigma_{\alpha \gamma}^2 &\equiv& \frac{F_2^*(\omega_\alpha,\omega_\alpha;\omega_\gamma,-\omega_\gamma)}{F_0^*(\omega_\alpha,\omega_\alpha)} \\
& \simeq& \frac{ \bigg{\la}  \bigg{\{} \left[ O(\omega_\alpha),H'(\omega_\gamma)\right] , \left[ H'(-\omega_\gamma), O^\dagger (\omega_\alpha)\right] \bigg{\}} \bigg{\ra}}{\bigg{\la}\bigg{\{ } O(\omega_\alpha), O^\dagger(\omega_\alpha) \bigg{\}}\bigg{\ra}}.~~~~~
\end{eqnarray}
Therefore, $\sigma_{\alpha \gamma}^2$  and $\sigma_{\alpha 0}^2$ will be positive and real in this approximation. The spectral function can be extracted according to $A(\omega) = -(1/\pi) Im G(\omega)$, with $G(\omega) = \int dt G(t) e^{i \omega t}$,
\begin{eqnarray}
\nonumber A^{(2)}(\omega) =&&\sum_{\alpha} F_0^*(\omega_\alpha,\omega_\alpha) \left( 1 - \sum_{\gamma\not=0} \frac{\sigma_{\alpha \gamma}^2}{\omega_\gamma^2}\right) \times\\
\nonumber && \hspace{0.5cm} \times \frac{1}{\sqrt{2\pi \sigma_{\alpha 0}^2}}e^{-\frac{[\omega-(\omega_\alpha - \sum_{\gamma\not=0}\frac{\sigma_{\alpha \gamma}^2}{\omega_\gamma})]^2}{2\sigma_{\alpha 0}^2}}+ \\
&&+\sum_\alpha \sum_{\gamma \not=0} \frac{F_0^*(\omega_\alpha, \omega_\alpha)}{\omega_\gamma^2}\delta\left(\omega -(\omega_\alpha + \omega_\gamma)\right).~~
\end{eqnarray}
Namely, the spectral function is the Gaussian line shifted by $\sum_{\gamma \not=0} \sigma_{\alpha \gamma}^2/\omega_\gamma$ from the original center $\omega_\alpha$ and broadened to width $\sigma_{\alpha 0}$. Besides, there is an additional resonance line, called a ``satellite line'' by Kubo and Tomita, with the relative intensity $\sigma_{\alpha \gamma}^2/\omega_\gamma^2$ at $\omega_\alpha + \omega_\gamma$.  Therefore, we can see that if the perturbation can be completely decomposed into Fourier series whose Fourier components do not develop time dependence, the perturbations broaden the original resonance line to a Gaussian line along with shifting the center, and there are additional resonance peaks (satellite lines). If the satellite lines are too close to the original resonance line peak, the perturbation will break down. Mathematically the perturbation breaks down if $\omega_\gamma \ll \sigma_{\alpha \gamma}$ or $\omega_\gamma \ll \sigma_{\alpha 0}$.

Now let's shift our focus on the second case in which the Fourier components of $H'(\omega_\alpha)$ develops time dependence. We will see that the Gaussian line will be modified to be a Lorentzian line, which was called motional narrowing by KT. In this case, the unperturbed Hamiltonian consists of two parts,$ H_0 = H_1 + H_2$, which satisfy the conditions
\begin{eqnarray}
&& \left[ H_1, H_2 \right] =0,\\
&& \left[ H_2, O\right] = 0.
\end{eqnarray}
$H_1$ is assumed to be not commutable with the operator $O$. With the conditions, we can straightforwardly adopt the approach from the first case. In the interaction picture, the unperturbed motion of the operator $O$ can be expanded as
\begin{eqnarray}
 \nonumber O^{(0)} = e^{iH_0 t} O e^{-i H_0 t} &=& e^{i H_1 t } O e^{-i H_1 t}= \sum_\alpha O(\omega_\alpha) e^{-i \omega_\alpha t},\\
\end{eqnarray}
where the Fourier component $O(\omega_\alpha)$ satisfy
\begin{eqnarray}
&& \left[ O(\omega_\alpha), H_1 \right] = \omega_\alpha O(\omega_\alpha),\\
&& \left[ O(\omega_\alpha),H_2 \right] = 0,
\end{eqnarray}
Similarly, we also introduce
\begin{eqnarray}
\nonumber H'(t) = e^{i H_0 t} H' e^{-i H_0 t} &=& e^{i H_2 t} e^{i H_1 t} H' e^{-i H_1 t} e^{-i H_2 t} \\
& = & \sum_\alpha H'(\omega_\alpha;t) e^{-i \omega_\alpha t},
\end{eqnarray}
where we introduce 
\begin{eqnarray}
&& e^{i H_1 t} H' e^{-i H_1 t} = \sum_\alpha H'(\omega_\alpha) e^{-i \omega_\alpha t},\\
&& e^{i H_2 t} H'(\omega_\alpha) e^{-i H_2 t} = H'(\omega_\alpha;t).
\end{eqnarray}
With the new definitions, we can adopt all the approaches illustrated in the first case with the replacement of $H'(\omega_\alpha)$ by the time-dependent $H'(\omega_\alpha;t)$. The general term of the expansion is
\begin{eqnarray}
\nonumber G_n(t) = && -i \theta(t) \sum_{\alpha, \beta} e^{-i \omega_\alpha t} \int_0^t dt_1 \cdots \int_0^{t_{n-1}}dt_n (i)^{-n} \times \\
\nonumber && \times \sum_{\gamma,\cdots, \nu} e^{-i(\omega_\gamma t_1 + \cdots + \omega_\nu t_n)} \times \\
\nonumber && \hspace{-1cm} \times \bigg{\la} \bigg{\{}\big{[} O(\omega_\alpha);H'(\omega_\gamma;t_1), \cdots, H'(\omega_\nu;t_n)\big{]},O^\dagger(\omega_\beta)\bigg{\}}\bigg{\ra}.\\
\end{eqnarray}
Again, due to the conservation of the frequency , we require
\begin{eqnarray}
\omega_\alpha - \omega_\beta + \omega_\gamma + \cdots +\omega_\nu =0.
\end{eqnarray}
The zero-th order term is 
\begin{eqnarray}
G_0(t) = -i \theta(t) \sum_\alpha e^{-i \omega_\alpha t} \bigg{\la}\bigg{\{} O(\omega_\alpha),O^\dagger(\omega_\alpha)\bigg{\}}\bigg{\ra}.
\end{eqnarray}
The first order term is
\begin{eqnarray}
\nonumber G_1(t) = &&-i \theta(t) \sum_{\alpha, \beta}\sum_{\gamma} \delta_{\omega_\alpha -\omega_\beta + \omega_\gamma=0}e^{-i \omega_\alpha t} \int_0^t dt_1 e^{-i\omega_\gamma t_1} \times \\
&& \hspace{1cm} \times \bigg{\la} \bigg{\{} \big{[} O(\omega_\alpha),H'(\omega_\gamma;t_1)\big{]},O^\dagger(\omega_\beta)\bigg{\}}\bigg{\ra}\\
\nonumber = && -i \theta(t) \sum_{\alpha, \beta}e^{-i \omega_\alpha t} \int_0^t dt_1 e^{i(\omega_\alpha -\omega_\beta) t_1} \times \\
&& \times \bigg{\la} \bigg{\{} \big{[} O(\omega_\alpha),H'(-\omega_\alpha + \omega_\beta; 0)\big{]},O^\dagger(\omega_\beta)\bigg{\}}\bigg{\ra},
\end{eqnarray}
which gives us the same result after integration to that in the first case, whose effects are to shift the center without broadening the resonance line. Let us focus on the second-order term, which is
\begin{eqnarray}
\nonumber && G_2(t) = -i \theta(t) \sum_{\alpha, \beta,\gamma,\delta}\delta_{\omega_\alpha -\omega_\beta + \omega_\gamma +\omega_\delta =0} e^{-i \omega_\alpha t} \times\\
\nonumber && \hspace{1.5 cm} \times \int_0^t dt_1 \int_0^{t_1} dt_2 e^{-i (\omega_\gamma t_1 + \omega_\delta t_2)}(i)^{-2} \times \\
\nonumber &&  \times \bigg{\la} \bigg{\{} \big{[}O(\omega_\alpha);H'(\omega_\gamma,t_1-t_2),H'(\omega_\delta,0)\big{]},O^\dagger(\omega_\beta)\bigg{\}}\bigg{\ra}.\\
\end{eqnarray}
Again, we focus on the contributions from $\omega_\alpha = \omega_\beta$, and $\omega_\gamma = - \omega_\delta$ and ignore highly asymmetric terms from $\omega_\beta \not=\omega_\beta$, which in most of the cases are not important. We get
\begin{eqnarray}
\nonumber && G_2(t) \simeq i \theta(t) \sum_{\alpha,\gamma} e^{-i \omega_\alpha t} \int_0^t dt_1 \int_0^{t_1} dt_2 e^{-i \omega_\gamma (t_1 - t_2)}\times\\
\nonumber && \times \bigg{\la} \bigg{\{} \big{[}O(\omega_\alpha);H'(\omega_\gamma,t_1-t_2),H'(-\omega_\gamma,0)\big{]},O^\dagger(\omega_\alpha)\bigg{\}}\bigg{\ra} \\
\nonumber &&= i \theta(t) \sum_{\alpha,\gamma} e^{-i \omega_\alpha t} \int_0^t d\tau(t-\tau)  e^{-i \omega_\gamma \tau}\times\\
\nonumber && \hspace{0.5 cm}\times \bigg{\la} \bigg{\{} \big{[}O(\omega_\alpha);H'(\omega_\gamma,\tau),H'(-\omega_\gamma,0)\big{]},O^\dagger(\omega_\alpha)\bigg{\}}\bigg{\ra}. \\
\end{eqnarray}
Following Kubo and Tomita, we introduce 
\begin{eqnarray}
\nonumber &&\sigma_{\alpha \gamma}^2 f_{\alpha \gamma}(\tau)  \\
\nonumber && = \frac{ \bigg{\la} \bigg{\{} \big{[}O(\omega_\alpha);H'(\omega_\gamma,\tau),H'(-\omega_\gamma,0)\big{]},O^\dagger(\omega_\alpha)\bigg{\}}\bigg{\ra}}{\bigg{\la} \bigg{\{}O(\omega_\alpha),O^\dagger(\omega_\alpha)\bigg{\}}\bigg{\ra}}\\
\nonumber && \simeq  \frac{ \bigg{\la} \bigg{\{} \big{[}O(\omega_\alpha),H'(\omega_\gamma,\tau)\big{]},\big{[}H'(-\omega_\gamma,0),O^\dagger(\omega_\alpha)\big{]}\bigg{\}}\bigg{\ra}}{\bigg{\la} \bigg{\{}O(\omega_\alpha),O^\dagger(\omega_\alpha)\bigg{\}}\bigg{\ra}}\\
\end{eqnarray}
where $\sigma_{\alpha \gamma}^2$ has been introduced before and $f_{\alpha \gamma}(\tau)$ is
\begin{eqnarray}
\nonumber && f_{\alpha \gamma}(\tau) \\
\nonumber && = \frac{\bigg{\la} \bigg{\{} \big{[}O(\omega_\alpha),H'(\omega_\gamma,\tau)\big{]},\big{[}H'(-\omega_\gamma,0),O^\dagger(\omega_\alpha)\big{]}\bigg{\}}\bigg{\ra}}{\bigg{\la} \bigg{\{} \big{[}O(\omega_\alpha),H'(\omega_\gamma,0)\big{]},\big{[}H'(-\omega_\gamma,0),O^\dagger(\omega_\alpha)\big{]}\bigg{\}}\bigg{\ra}},\\
\end{eqnarray}
which is reduced to be $1$ if $H'(\omega_\gamma,\tau)$ becomes time independent. Then the expression for the second ordre term is
\begin{eqnarray}
\nonumber G_2(t) \simeq && i \theta(t) \sum_{\alpha, \gamma}e^{-i\omega_\alpha t} \bigg{\la} \bigg{\{}O(\omega_\alpha),O^\dagger(\omega_\alpha)\bigg{\}}\bigg{\ra} \times\\
&& \times \int_0^t d\tau (t-\tau) e^{-i \omega_\gamma \tau} \sigma_{\alpha \gamma}^2 f_{\alpha \gamma}(\tau).
\end{eqnarray} 
Now, let us combine the terms up to second order. Since first order term is the same to that in the first case, it will simply shift the resonance line, so we ignore it. Up to second order, we get
\begin{eqnarray}
\nonumber G^{(2)} (t) &\simeq& G_0(t) + G_2(t) \\
\nonumber & \simeq & G_0(t) \sum_{\gamma} \bigg{(} 1 - \int_0^t d\tau (t-\tau) \sigma_{\alpha \gamma}^2 f_{\alpha \gamma}(\tau)\bigg{)} \\
&\simeq & G_0(t)\sum_\gamma e^{-\int_0^t d\tau (t-\tau) \sigma_{\alpha \gamma}^2 f_{\alpha \gamma}(\tau)}.
\end{eqnarray}
Focusing on the simplest case with $\gamma =0$, we have
\begin{eqnarray}
G^{(2)} (t) \simeq G_0(t) e^{-\sigma_{\alpha 0}^2 \int_0^t d\tau (t-\tau)f_{\alpha 0}(\tau)}.
\end{eqnarray}
At the long time limit, $t\rightarrow \infty$, we can approximate
\begin{eqnarray}
G^{(2)}(t\rightarrow \infty) &\sim& G_0(t) e^{-\sigma_{\alpha 0}^2 \int_0^\infty d\tau t f_{\alpha 0}(\tau)} \\
& = & G_0(t) e^{- \sigma_{\alpha 0}^2  t \left( \tau' + i\tau'' \right)},
\end{eqnarray}
where we introduce $\tau'\equiv \int_0^\infty Re\{ f_{\alpha 0}(\tau) \} d\tau $ and $\tau'' \equiv \int_0^\infty Im\{ f_{\alpha 0}(\tau) \} d\tau$. Then the spectral function becomes
\begin{eqnarray}
A^{(2)}(\omega) \simeq \frac{1}{\pi} \frac{\sigma_{\alpha }^2 \tau'}{[\omega- (\omega_\alpha -\sigma_{\alpha 0}^2 \tau'')]^2 + (\sigma_{\alpha 0}^2 \tau')^2},
\end{eqnarray}
where we can see that the line shape becomes Lorentzian! Note that if we set $f_{\alpha 0}(\tau) = \bar{f}_{\alpha 0}$, independent of time, the shape will go back to Gaussian. In general, if we can get the ``exact'' form of $f_{\alpha \gamma}(\tau)$, we can obtain the exact form of the resonance line, which is, unfortunate, not possible in most cases.

\section{Perturbative studies in the $U=0$ limit}\label{APP:perturbation_U=0}
\begin{figure}[t]
   \centering
   \includegraphics[width=2.5 in]{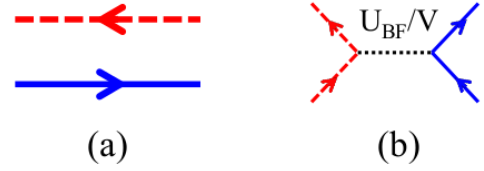}
   \caption{(Color Online) (a) The Goldstino Green's function consisting of a fermion Green's function (red dashed line) and a boson Green's function (blue solid line) whose momenta are explicitly summed. (b) The two-fermion-two-boson vertex arising from $H' = U_{BF} \sum_j \hat{n}^b_j \hat{n}^f_j = \frac{U_{BF}}{V} \sum_{{\bf p}, {\bf q}, {\bf k}} f^\dagger_{\bf p} f_{{\bf p} + {\bf q} - {\bf k}}b^\dagger_{\bf q} b_{\bf k}$.}
   \label{Fig:Goldstino-FD}
\end{figure}
\begin{figure}[t]
   \centering
   \includegraphics[width=3 in]{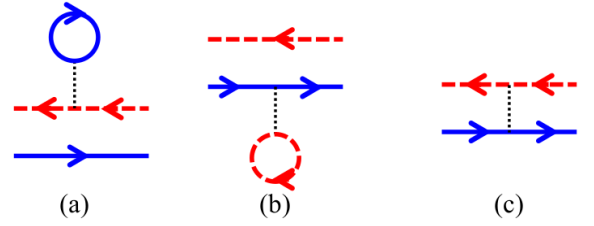}
   \caption{(Color Online) First-order Feynman diagrams contributing to renormalize the Goldstino Green's function. (a) A boson bubble in the fermion Green's function sector renormalizing $\mu_f$. (b) A fermion buble in the boson Green's function sector renormalizing $\mu_b$. (c) A two-fermi-two-bose vertex contributing to the first-order self-energy.}
   \label{Fig:1st-FD}
\end{figure}
\begin{figure}[t]
   \centering
   \includegraphics[width=3 in]{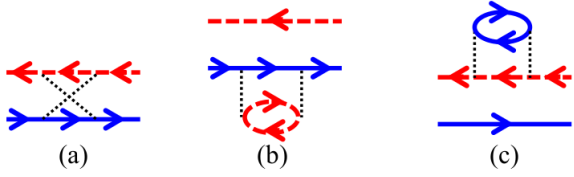}
   \caption{(Color Online) The second-order irreducible Feynman diagrams. (a) Two cross interaction vertices. (b) A fermionic particle-hole bubble in the boson Green's function sector. (c) A bosonic particle-hole bubble in the fermion Green's function sector. }
   \label{Fig:2nd-FD}
\end{figure}
We present the perturbative studies of the broadening of the spectral function of Goldstino mode at $U=0$. The Hamiltonian is $\hat{H} = \hat{H}_0 + \hat{H}'$ with
\begin{eqnarray}
&& \hat{H}_0 =  - t_h \sum_{\la j k \ra} \left[ b^\dagger_j b_k + f^\dagger_j f_k\right] - \mu_f \hat{N}_f - \mu_b \hat{N}_b, \\
&&\hat{H}' =U_{BF} \sum_j \hat{n}_j^b \hat{n}_j^f.
\end{eqnarray}
For convenience, we introduce the imaginary-time-ordered Goldstino Green's function 
\begin{eqnarray}
\mathcal{G}(\tau) \equiv - \left\la T_\tau \left[ Q(\tau) Q^\dagger(0)\right]\right\ra,
\end{eqnarray}
where we remind that the Goldstino field is
\begin{eqnarray}
Q\equiv \frac{1}{\sqrt{N}}\sum_j f^\dagger_j b_j.
\end{eqnarray}
The general expression for the perturbation is
\begin{eqnarray}
\nonumber \mathcal{G}(\tau) =&& - \sum_{m=0}^\infty  \frac{(-1)^m}{m!} \int_0^\beta d\tau_1 \cdots \int_0^\beta d\tau_m \times \\
&& \times \left\la T_\tau\left[ \hat{H}'(\tau_1)\cdots \hat{H}'(\tau_m)Q(\tau)Q^\dagger(0)\right]\right\ra_{0, conn},
\end{eqnarray}
where the subscript $0$ means the unperturbed ground state and $conn$ means the connected Feynman diagrams.

At zeroth order, we know that
\begin{eqnarray}
\nonumber \mathcal{G}^{(0)}(\tau) &=& - \left\la T_\tau \left[ Q(\tau) Q^\dagger(0)\right]\right\ra \\
\nonumber &=& - \frac{1}{N}\sum_{{\bf p},{\bf q}} \left\la T_\tau \left[ f^\dagger_q(\tau)b_q(\tau)f_p(0)b^\dagger_p(0)\right]\right\ra\\
\nonumber &=&\frac{1}{N} \sum_{{\bf p},{\bf q}}\left\la T_\tau \left[ f_p(0)f^\dagger_q(\tau)\right]\right\ra \left\la T_\tau\left[b_q(\tau)b_p^\dagger(0)\right]\right\ra\\
&=&\frac{1}{N}\sum_{\bf p} \mathcal{G}_f(-\tau,{\bf p})\mathcal{G}_b(\tau,{\bf p}),
\end{eqnarray}
where $\mathcal{G}_{f/b}(\tau,{\bf p})$ are the imaginary-time ordered fermion/boson Green's functions, and we know
\begin{eqnarray}
&& \mathcal{G}_f(-\tau,{\bf p}) = \left[ \theta(\tau) n_f(\epsilon_{\bf p}^f) -\theta(-\tau)\left(1-n_f(\epsilon_{\bf p}^f) \right) \right] e^{\epsilon^f_{\bf p}\tau},\\
&& \mathcal{G}_b(\tau,{\bf p})= - \left[ \theta(\tau)\left(1+n_b(\epsilon_{\bf p}^b)\right) + \theta(-\tau)n_b(\epsilon_{\bf p}^b \right]e^{-\epsilon_{\bf p}^b \tau},~~~~~~~~
\end{eqnarray}
where we introduce $n_{f/b}(z) \equiv ( e^{\beta z} \pm1)^{-1}$ and $\epsilon_{\bf p}^{f/b} \equiv \omega_{\bf p}^{f/b} - \mu_{f/b}$, with $\omega_{\bf p}^{f/b}$ being the dispersions of hopping. Introducing the Fourier transform $\mathcal{G}(i\omega_n) = \int_0^\beta d\tau \mathcal{G}(\tau)e^{i \omega_n \tau}$ with $\omega_n = (2n+1)\pi/\beta (2n\pi/\beta)$ for fermions/bosons, we get
\begin{eqnarray}
\mathcal{G}^{(0)}(i\omega_n) = \frac{1}{N}\sum_{\bf p} \frac{ n_f(\epsilon_{\bf p}^f)+ n_b (\epsilon_{\bf p}^b)} {i\omega_n + \epsilon_{\bf p}^f - \epsilon_{\bf p}^b}= \frac{1}{i\omega_n - \Delta\mu},~~
\end{eqnarray}
where we note that $\omega_n = (2n+1)\pi/\beta$ since the Goldstino field is a fermion. The zeroth order (unperturbed) retarded Goldstino Green's function can be obtained as
\begin{eqnarray}
G^{R,(0)}(\omega)=\mathcal{G}(i\omega_n \rightarrow \omega + i 0^+)= \frac{1}{\omega -\Delta\mu+i 0^+},~~~
\end{eqnarray}
which is the same to that derived from the equation of motion presented in the main texts. For clarity in presenting the perturbative studies below, we introduce the diagram presentation for the Goldstino Green's function in Fig.~\ref{Fig:Goldstino-FD}(a), where the upper dashed red line represents the fermion line and the bottom blue solid line represents the boson line and the momenta of fermion and boson lines are explicitly summed. Fig.~\ref{Fig:Goldstino-FD}(b) represents the two-fermion-two-boson vertex. The arrows represent the directions of the particles.  

For the first order term, we find there are three possible contributions as shown in Fig.~\ref{Fig:1st-FD}. The first-order perturbation gives (in imaginary time domain)
\begin{eqnarray}
\nonumber &&\mathcal{G}_1(\tau) = \frac{U_{BF}}{N}\int_0^\beta d\tau_1 \times \\
&& \times \bigg{[}  \rho_B \sum_{\bf p} \mathcal{G}_f(\tau_1 - \tau,{\bf p})\mathcal{G}_f(-\tau_1,{\bf p}) \mathcal{G}_b(\tau,{\bf p}) +\\
\nonumber && \hspace{0.5 cm} + \rho_F \sum_{\bf p} \mathcal{G}_f(-\tau,p) \mathcal{G}_b(\tau_1,p)\mathcal{G}_b(\tau-\tau_1, {\bf p}) - \\
\nonumber && - \frac{1}{V} \sum_{{\bf p},{\bf q}} \mathcal{G}_f(\tau_1-\tau,{\bf q})\mathcal{G}_b(\tau-\tau_1,{\bf q}) \mathcal{G}_f(-\tau_1,{\bf p}) \mathcal{G}_b(\tau_1,{\bf p})\bigg{]}.\\
\end{eqnarray}
Going to Fourier space, we get
\begin{eqnarray}
\nonumber \mathcal{G}_1(i\omega_n) &=& \mathcal{G}^{(0)}(i\omega_n)\left( - \rho_B U_{BF} + \rho_F U_{BF} -\rho U_{BF}\right) \mathcal{G}^{(0)}(i\omega_n) \\
&=& \mathcal{G}^{(0)}(i \omega_n) \left( -2 \rho_B U_{BF}\right) \mathcal{G}^{(0)}(i\omega_n).
\end{eqnarray}
Therefore, we can see if we only include the first order term, the renormalized imaginary-time Goldstino Green's function at first order is
\begin{eqnarray}
\mathcal{G}^{(1)}(i\omega_n) \simeq \frac{1}{i\omega_n - \left(\Delta\mu - 2 \rho_B U_{BF}\right)},
\end{eqnarray}
which gives
\begin{eqnarray}
G^{R,(1)}(\omega) = \frac{1}{\omega - \left(\Delta \mu - 2 \rho_B U_{BF} \right) + i 0^+},
\end{eqnarray}
which again is consistent with the result obtained by Kubo-Tomita theory presented in the main texts and leads to the spectral function
\begin{eqnarray}
A^{(1)}(\omega) = \delta (\omega - (\Delta \mu - 2\rho_B U_{BF})).
\end{eqnarray}
For the second order, we find there are three contributions illustrated schematically in Fig.~\ref{Fig:2nd-FD}. The derivation for the second-order terms are straightforward (but tedious), and we present the final results below. Fig~\ref{Fig:2nd-FD}(a)-(c) contributing to renormalize the imaginary-time Goldstino Green's functions can be concisely written as $\mathcal{G}^{(0)}\left( \Sigma^{(2)}_1 + \Sigma^{(2)}_2 + \Sigma^{(2)}_3 \right)\mathcal{G}^{(0)}$, where the subscript $(2)$ means the second-order and the subscripts $1-3$ label the contributions from Fig.~\ref{Fig:2nd-FD}(a)-(c).  The self-energies $\Sigma$ contributed from each term are
\begin{widetext}
\begin{eqnarray}
\Sigma^{(2)}_1 (i\omega_n)= \frac{U_{BF}^2}{N V^2} \nonumber \sum_{{\bf p},{\bf q},{\bf k}} \bigg{\{} &&\frac{\left[ n_f(\epsilon_{\bf p}^f) + n_b(\epsilon_{\bf k}^b)\right]\left[ n_f(\epsilon_{{\bf p} + {\bf q} - {\bf k}}^f) - n_f(\epsilon_{\bf p}^f - \epsilon_{\bf k}^b)\right]\left[ n_f(\epsilon_{\bf q}^f) + n_b(-\epsilon_{\bf p}^f + \epsilon_{\bf k}^b + \epsilon_{{\bf p} + {\bf q} - {\bf k}}^f)\right]}{i\omega_n + \epsilon_{\bf p}^f + \epsilon_{\bf q}^f - \epsilon_{\bf k}^b - \epsilon_{{\bf p} + {\bf q} - {\bf k}}^f} \\
\nonumber &&+ \frac{\left[ n_b(\epsilon_{\bf k}^b) - n_b(\epsilon_{\bf p}^b)\right]\left[ n_f(\epsilon_{{\bf p} + {\bf q} -{\bf k}}^f) + n_b(\epsilon_{\bf p}^b - \epsilon_{\bf k}^b)\right]\left[ n_b(\epsilon_{\bf q}^b) + n_f(-\epsilon_{\bf p}^b + \epsilon_{\bf k}^b - \epsilon_{{\bf p} + {\bf q} - {\bf k}}^f)\right]}{i\omega_n - \epsilon_{\bf p}^b - \epsilon_{\bf q}^b + \epsilon_{\bf k}^b + \epsilon_{{\bf p} + {\bf q} - {\bf k}}^f}\\
\nonumber &&+ \frac{\left[ n_b(\epsilon_{\bf p}^b - n_b(\epsilon_{\bf k}^b)\right]\left[n_f(\epsilon_{{\bf p} + {\bf q} -{\bf k}}^f) + n_b(\epsilon_{\bf p}^b - \epsilon_{\bf k}^b)\right]\left[n_f(\epsilon_{\bf q}^f)-n_f(-\epsilon_{\bf p}^b+\epsilon_{\bf k}^b+\epsilon_{{\bf p}+{\bf q} - {\bf k}}^f)\right]}{\epsilon_{\bf p}^b + \epsilon_{\bf q}^f - \epsilon_{\bf k}^b - \epsilon_{{\bf p}+ {\bf q} - {\bf k}}^f} \\
&& + \frac{\left[n_f(\epsilon_{\bf p}^f) + n_b(\epsilon_{\bf k}^b)\right]\left[ n_f(\epsilon_{{\bf q} + {\bf q} -{\bf k}}^f) - n_f(\epsilon_{\bf p}^f - \epsilon_{\bf k}^b)\right]\left[n_b(\epsilon_{\bf q}^b) -n_b(-\epsilon_{\bf p}^f + \epsilon_{\bf k}^b + \epsilon_{{\bf p} + {\bf q} - {\bf k}}^f)\right]}{\epsilon_{\bf p}^f + \epsilon_{\bf q}^b - \epsilon_{\bf k}^b - \epsilon_{{\bf p} + {\bf q} -{\bf k}}^f}\bigg{\}},~~
\end{eqnarray}
\begin{eqnarray}
\nonumber \Sigma^{(2)}_2(i\omega_n) = \frac{U_{BF}^2}{NV^2}\sum_{{\bf p}, {\bf q},{\bf k}}\bigg{\{} &&\frac{- n_f(\epsilon_{\bf p}^f) \left[ n_f(\epsilon_{{\bf q} + {\bf k}-{\bf p}}^f) - n_f(\epsilon_{\bf k}^b-\epsilon_{\bf p}^b)\right]\left[ n_f(\epsilon_{\bf q}^f) + n_b(-\epsilon_{\bf k}^f + \epsilon_{\bf p}^b+ \epsilon_{{\bf q} + {\bf k} -{\bf p}}^f)\right]}{i\omega_n - \epsilon_{\bf q}^f - \epsilon_{\bf k}^b + \epsilon_{\bf p}^b + \epsilon_{{\bf q} + {\bf k} - {\bf p}}^f} \\
&&+ \frac{n_b(\epsilon_{\bf p}^f)\left[ n_f(\epsilon_{{\bf q} + {\bf k} - {\bf p}}^f) + n_b( \epsilon_{\bf k}^b-\epsilon_{\bf p}^b)\right]\left[ n_f(\epsilon_{\bf q}^f) - n_f( -\epsilon_{\bf k}^b + \epsilon_{\bf p}^b + \epsilon_{{\bf q} + {\bf k} - {\bf p}}^f)\right]}{\epsilon_{\bf q}^f + \epsilon_{\bf k}^b - \epsilon_{\bf p}^b - \epsilon_{{\bf q} + {\bf k} -{\bf p}}^f}\bigg{\}},
\end{eqnarray}
\begin{eqnarray}
\nonumber \Sigma^{(2)}_3 (i\omega_n)= \frac{U_{BF}^2}{NV^2} \sum_{{\bf p},{\bf q}, {\bf k}} \bigg{\{} && \frac{n_b(\epsilon_{\bf p}^b)\left[ n_f(\epsilon_{\bf q}^f) + n_b(\epsilon_{\bf p}^b + \epsilon_{{\bf q} + {\bf k} - {\bf p}}^b)\right]\left[n_b(\epsilon_{\bf k}^b) + n_f(-\epsilon_{\bf q}^f + \epsilon_{\bf p}^b + \epsilon_{{\bf q} + {\bf k} -{\bf p}}^b)\right]}{i\omega_n + \epsilon_{\bf q}^f + \epsilon_{\bf k}^b -\epsilon_{\bf p}^b - \epsilon_{{\bf q} + {\bf k} -{\bf p}}^b} \\
&& + \frac{n_f(\epsilon_{\bf p}^f)\left[ n_f (\epsilon_{\bf q}^f) - n_f(\epsilon_{\bf p}^f + \epsilon_{{\bf q} + {\bf k} -{\bf p}}^b)\right]\left[ n_b(\epsilon_{\bf k}^b) - n_b(-\epsilon_{\bf q}^f + \epsilon_{\bf p}^f + \epsilon_{{\bf q} + {\bf k} - {\bf p}}^b)\right]}{\epsilon_{\bf q}^f + \epsilon_{\bf k}^b - \epsilon_{\bf p}^f - \epsilon_{{\bf q} + {\bf k} - {\bf p}}^b}\bigg{\}}.
\end{eqnarray}
\end{widetext}
The second-order self-energies in the Matsubara frequency domain are very complicated. However, since we are only interested in the \textit{imaginary} part of the self-energy in the real frequency domain obtained by $i\omega_n \rightarrow \omega + i 0^+$ and extract its qualitative behavior. In this way, we can see that $\Sigma^{(2)}_{a=1,2,3}(\omega)$ all contains imaginary parts which are \textit{proportional to } square of the perturbation strength, i. e. $Im \Sigma^{(2)} (\omega) \equiv \Gamma^{(2)} \propto U_{BF}^2$. In the presence of the imaginary part of the self-energy, the retarded Goldstino Green's function becomes (we ignore the real part of $\Sigma^{(2)(\omega)}$ since it only contributes to shift the location of the peak)
\begin{eqnarray}
\mathcal{G}^{R,(2)}(\omega) \simeq \frac{1}{\omega -(\Delta \mu - 2 \rho_B U_{BF}) - i \Gamma^{(2)}},
\end{eqnarray}
which leads to the spectral function
\begin{eqnarray}
A^{(2)}(\omega) \simeq \frac{\Gamma^{(2)}}{(\omega - (\Delta \mu - 2 \rho_B U_{BF}))^2 + (\Gamma^{(2)})^2},
\end{eqnarray}
which is a Lorentzian function with width $\sim \Gamma ^{(2)} \propto U_{BF}^2$. Therefore, we can see that the perturbative results at $U=0$ limit are qualitatively consistent with the results obtained by Kubo-Tomita theory.
\bibliography{biblio4Goldstino}
\end{document}